\documentclass[12pt]{iopart}
\usepackage{iopams}
\usepackage{graphicx}
\usepackage{times,txfonts}
\usepackage{amsfonts}
\usepackage{subfig}
\begin{document}
\title{Unveiling the Hanbury Brown and Twiss effect through R\'enyi entropy correlations}
\author{Sammy Ragy and Gerardo Adesso}
\address{School of Mathematical Sciences, The University of Nottingham,
University Park, Nottingham NG7 2RD, United Kingdom}
\begin{abstract} Adopting a quantum information perspective, we analyse the correlations in the thermal light beams used to demonstrate the Hanbury Brown and Twiss effect. We find that the total correlations measured by the R\'enyi  mutual information match the normalised intensity correlations in the regime of low source intensity. Genuine quantum correlations in the form of  discord are relevant in such regime but get washed out with increasing source intensity. This provides a new angle on the issue about the nature---quantum versus classical---of the effect.
\end{abstract}
\section{Introduction}

\begin{figure}[ht]
\centering
\includegraphics[width=7cm]{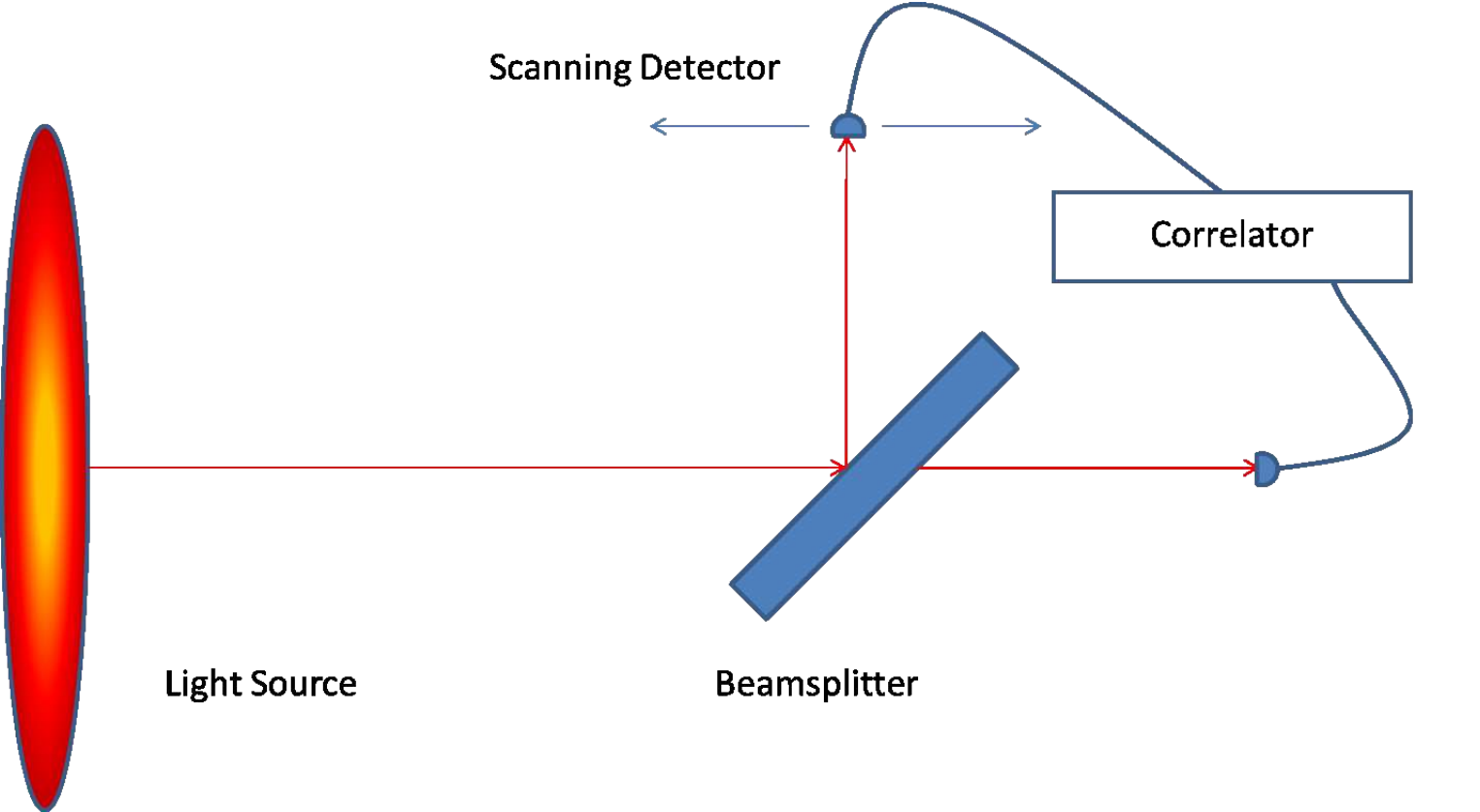}
\caption{\footnotesize{Diagrammatic representation of a Hanbury Brown and Twiss set-up. Light from a large source propagates into the far field whereupon it is split on a balanced beamsplitter. We denote by $\hat{a}$ the mode operators of the source, by $\hat{a}_{vac}$ the mode operators of the vacuum field entering the other arm of the beamsplitter, and by $\hat{b}_{1,2}$ the outcoming mode operators in the two arms after the beamsplitter. When one of the detectors is scanned in the transverse plane, the normalised intensity correlations of the modes $\hat{b}_{1,2}$  exhibit an oscillatory behaviour as a function of the transverse coordinate $x$ of the scanning detector. We find that these intensity correlations are matched by the total quadrature correlations (as measured by R\'enyi-$2$ mutual information) in the bipartite system formed by the modes $\hat{b}_{1,2}$ in the limit of low source photon-count. In that regime, a relevant portion of such total correlations is constituted by genuinely quantum ones as measured by discord.}}
\label{scheme}
\end{figure}

The Hanbury Brown and Twiss (HBT) effect was demonstrated in 1956 \cite{HBT1} and had far-reaching impact, even inspiring some of Glauber's seminal work on quantum optics \cite{Glauber}. It was the first demonstration of the second-order correlation of visible light and caused significant debate; some held the belief that it ran counter to the photon theory of light and the experimental results were therefore incorrect \cite{Brannen, Purcell, HBT2}. After a series of papers by Hanbury Brown and Twiss, the controversy was largely abated \cite{HBT4, HBT5}. However, even in the past few years, papers have been published which argue \textit{against} the established interpretation \cite{Scarcelli, Shih}.

This revived interest comes about from an analogy between the HBT effect and the phenomenon of ghost imaging: simply put, both phenomena depend upon intensity correlations. Due to questions surrounding the classical interpretations of ghost imaging, similar suspicions were raised upon the HBT effect by Scarcelli et al. who suggested that its explanation relies on non-local two photon interference \cite{Scarcelli}, although this claim has proved controversial \cite{GIHBT}.

The HBT effect can be demonstrated by a set-up as shown in figure 1. Light from a distant source impinges upon a balanced beamsplitter and then it propagates down two separate, yet identical, arms of the configuration. Subsequently, each beam falls upon a point-like detector. When one detector is kept fixed and the other detector is scanned in the transverse plane, we observe a correlation of the intensity in what is known as the HBT effect.

Prior to the 1956 HBT paper, such effects had been demonstrated in radio astronomy and were readily accepted as permitting a simple, classical explanation \cite{HBT3}. This is because the experiments had all been performed in the regime of large photon-number per mode, in which the classical theory of light easily holds sway \cite{Fellgett} and the shot noise does not contribute a large extent. In this case, we can regard the beamsplitter as creating two identical duplicates of the intensity profile of light. Then by the van Cittert-Zernike theorem it is possible to classically calculate the correlation function as we scan one detector across the beam \cite{Mandel}. Essentially, the light propagates in such a manner that the classical intensity fluctuations are appreciably correlated at the detection plane over a finite area.

However, the 1956 experiment involved measuring the correlations of optical photons from a mercury arc lamp, in which each mode contains few photons. In this case, it becomes more difficult to understand how it can be that some sort of `classical' behaviour emerges even when few photons impinge upon the detectors. However, it was proven by Purcell \cite{Purcell}, that the experimental results of HBT could be explained by using a classical light field as long as the electrons in the detector were given a quantum treatment.

The success of this `semi-classical' model later came to be understood through the generalised formalism of quasi-probability distributions \cite{Mandel}. In 1963, during the birth of quantum optics in its current form, Sudarshan proved that any experiments involving light which is characterised by a positive, well-behaved $P$-representation can also be fully explained by semi-classical methods \cite{Sudarshan}. Thermal light -- such as that present in the HBT experiment -- satisfies this constraint.

The notion of classicality which emerges from the $P$-representation (P-classicality), however, is by no means the only notion available. In quantum information, a popular and alternative definition of classicality for bipartite states arises from the concept of `discord', a quantity which is defined as the difference between two classically identical expressions for mutual information \cite{Zurek,Vedral}. The discord can be considered to represent the degree of quantum correlations between each part of the bipartition and when the discord is equal to zero, this defines an indication of classicality (D-classicality) of the correlations in the state. It has recently been shown that both notions of classicality are maximally inequivalent \cite{Paris}.

In this paper we follow the example of a recent treatment of ghost imaging in terms of discord \cite{Ourpaper}. Since the HBT effect is entirely contingent upon the correlations between each arm, we believe that the very general informational approach to correlations as defined by the mutual information (total correlations) and discord (quantum correlations) provides an apt foundation for studying (non)classicality within the scheme. In particular, we exploit a powerful new variant of mutual information (and its associated classical and quantum correlations) defined using the R\'enyi-$2$ entropy for Gaussian states \cite{Renyi}. This definition of mutual information is intimately related to phase space sampling by homodyne detections which serves as a strong interpretational aid as well as harbouring numerous other desirable properties \cite{Renyi}.

We establish that, despite the availability of a semi-classical light model, the quantum correlations are indeed significant in the regime of low photon-count per mode in which the original controversy was conceived. Thus, whilst P-classicality holds, we find an undeniable quantum signature as revealed by a relevant portion of quantum correlations (discord). This is somehow reflected in the physical model proposed by Scarcelli et al. \cite{Scarcelli} in which non-local correlations feature in the imaging. We further find that in the limit of a very low photon-count per mode, the mutual information tends to coincide analytically with the normalised intensity correlations.

\section{Preliminaries}
\subsection{R\'enyi entropy information measures for Gaussian states}
A detailed account of the correlations and information measures in terms of R\'enyi-$2$ entropy can be found in \cite{Renyi}. To summarise, the definitions reflect the usual formulation of mutual information, classical correlations and quantum discord \cite{Zurek, Vedral} but with the replacement of the von Neumann entropy with the R\'enyi-$2$ entropy.

A R\'enyi-$\alpha$ entropy is defined by
\begin{eqnarray}
S_\alpha=(1-\alpha)^{-1} \ln \tr (\rho^\alpha)
\end{eqnarray}
The von Neumann entropy which is used as a conventional measure of quantum information can be obtained by taking the limit $\alpha \rightarrow 1$, whereas for the purposes of this paper we use the R\'enyi-$2$ entropy, which is defined by $\alpha=2$, and simply reduces to minus the logarithm of the purity of a quantum state.

For this choice, and given a bipartite Gaussian state $\rho_{AB}$ with marginal states $\rho_{A}$ ($\rho_B$) for subsystems $A$ ($B$),  the R\'enyi-$2$ mutual information, classical correlations and quantum discord are defined respectively by
\begin{eqnarray}
\mathcal{I}(A:B)&= S_{2}(\rho_A)+S_{2}(\rho_B)-S_2(\rho_{AB}) \label{eqI}\\
\mathcal{J}(A|B)&= S_{2}(\rho_A)-\inf_{\Pi_i}\mathcal{H}_2(A|B_{\Pi_i}) \label{eqJ}\\
\mathcal{D}(A|B)&= S_{2}(\rho_B)-S_{2}(\rho_{AB})+\inf_{\Pi_i}\mathcal{H}_2(A|B_{\Pi_i}) \label{eqD}
\end{eqnarray}
Here, we have introduced a quantum conditional entropy $\mathcal{H}_2(A|B_{\Pi_i})=\sum_i p_i S_2(\rho_{A|i})$ which quantifies the entropy on $A$ after a measurement $\{\Pi_i\}$ has been performed on $B$. Herein lies the subtlety which distinguishes the quantum from classical case. In the classical case, the conditional entropy factors out, and the discord evaluates to $0$. In the quantum case, however, we need to consider the measurement performed when calculating the conditional entropy. Since a measurement can disturb the system, the means we cannot fully extract all the information of a quantum system. Taking the \textit{least} disturbing measurement, and restricting this minimisation to Gaussian measurements, we can calculate the Gaussian classical correlations  $\mathcal{J}$ and quantum discord  $\mathcal{D}$ as defined in the above equations (\ref{eqJ}-\ref{eqD}) \cite{QvC,Giorda}.

We remind that the mutual information $\mathcal{I}$ defined in terms of the R\'enyi-$2$ entropy as in equation (\ref{eqI}), of a particular Gaussian state $\rho_{AB}$, corresponds to the Shannon continuous mutual information of its Wigner distribution. More generally, it stems from the satisfaction of the strong subadditivity inequality by the R\'enyi-$2$ entropy (on Gaussian states), that it can be used to reformulate the core of quantum information theory and to define valid correlation measures in the Gaussian setting \cite{Renyi}.

\subsection{Covariance matrix description of the HBT effect}
In order to compute the discord and other correlations for Gaussian states, it is necessary to produce the covariance matrix of the relevant quadrature operators \cite{QvC}. Such a matrix fully characterises the state up to local displacements. For a general two-mode Gaussian state $\rho_{AB}$, any covariance matrix  $\boldsymbol{\sigma}_{AB}$ can be written (by means of local unitaries) in the standard form,
\begin{eqnarray}
\boldsymbol{\sigma}_{AB}=\left(\begin{array}{cccc}
a & 0 &c & 0 \\
0 & a & 0 & d\\
c&0 & b & 0 \\
0& d &0 &b
\end{array} \right).
\label{Covmat}
\end{eqnarray}

For the HBT setting under consideration (see Fig.~\ref{scheme}), the entries are easily calculated with some simplifying assumptions: firstly, we choose a large disk-like monochromatic source of area $A$, emitting spatially incoherent light. From this, we take each mode $\hat{a}(q)$ (where $q$ is the transverse component of the wave-vector) to be uncorrelated with every other mode in the source plane, and we consider all modes to be zero-mean modes with identical photon-number expectation value $\bar{n}$, such that $\langle\hat{a}^\dagger(q) \hat{a}(q') \rangle= \bar{n} \delta(q-q')$, these are common assumptions for intensity interferometric experiments with thermal light \cite{Scarcelli,Gatti}.
This implies that in the source plane
$\langle\hat{a}^\dagger(x_2) \hat{a}(x_1) \rangle=\int e^{-i\mathbf{q}\cdot\mathbf{x_1}}e^{i\mathbf{q}'\cdot\mathbf{x_2}}\langle\hat{a}^\dagger(q') \hat{a}(q) \rangle dq dq'=\bar{n} \delta(x_1-x_2)$,
where $x_i$ represent transverse position vectors.
Furthermore, we take the beamsplitter to be balanced (50:50) such that the outcoming mode operators for each branch are
\begin{eqnarray}
\hat{b}_{1,2}(x)=\frac{1}{\sqrt{2}}(\hat{a}(x) \pm \hat{a}_{vac}(x)).
\end{eqnarray}
Now we can calculate the covariance between all four mode operators of relevance $\hat{b}_i$ and $\hat{b}_i^\dagger$, with $i=1,2$ (after which we can rotate into the quadrature basis). To do this, we also need to take into account the propagation of the incoming modes $\hat{a}(x)$ through space. We consider the Fraunhofer propagator $g(x_1,x)=e^{-i\frac{k}{z}(x_1\cdot x)}$ which relates the modes in the source plane to the observation plane by
$\hat{a}(x)\propto \int_{source} g(x_1,x) \hat{a}(x_1) dx_1.$
As an example, we can then calculate and report here the correlation function $\langle\hat{a}^\dagger(x) \hat{a}(x') \rangle$,
\begin{eqnarray}
\langle\hat{a}^\dagger(x) \hat{a}(x') \rangle &\propto  \int_{A}\int_{A} g(x_1,x) g^*(x_2,x') \langle \hat{a}(x_1) \hat{a}(x_2)\rangle d^2 x_1 d^2x_2 \nonumber  \\
&= \int_{A}\int_{A} e^{i\frac{k}{z}(x_2\cdot x'-x_1\cdot x)} \bar{n} \delta(x_1-x_2)d^2x_1 d^2x_2 \nonumber \\
&=\bar{n} \int_{A} e^{-i\frac{k}{z}(x_1\cdot(x-x'))}dx^2_1 \nonumber  \\
&=\bar{n} \frac{2J(\frac{kA}{z}|x-x'|)}{\frac{kA}{z}|x-x'|} =\bar{n} \,\mbox{Jinc}\left(\frac{kA}{z}|x-x'|\right)\,,
\end{eqnarray}
where $J$ refers to a Bessel function of the first kind and we have defined $\mbox{Jinc}(y)=\frac{2J(y)}{y}$. Continuing along these lines and setting $x'=0$ we obtain a covariance matrix $\boldsymbol{\sigma}_{12}(x)$ in standard form as in equation~(\ref{Covmat}), where \begin{eqnarray}
a=b&=&1+2\bar{n}\,, \quad \hbox{and} \\
c=d&=&2\bar{n} \, \mbox{Jinc}\left(\frac{kA}{z}|x|\right)\,. \nonumber
\label{cmelements}\end{eqnarray}

\subsection{Measures of correlations}
From the previous analysis, and using the results of \cite{Renyi}, we can obtain the explicit expressions for the R\'enyi-$2$ mutual information, classical correlations and discord as functions of $x$,
\begin{eqnarray}
\mathcal{I}(A:B)&= \ln\left[\frac{a^2}{a^2-c^2}\right],\\
\mathcal{J}(A|B)&= \ln\left[\frac{a^2+a}{a^2+a-c^2}\right],\\
\mathcal{D}(A|B)&= \ln\left[\frac{a^2+a^3-ac^2}{a^2+a^3-ac^2-c^2}\right].
\end{eqnarray}

We find it relevant  at this point to recall also the formula for the normalised intensity correlation function in our scheme. Defining the intensity in each beam $i=1,2$ as $\hat{I}_i(x)=\hat{b}^\dagger_i(x)\hat{b}_i(x)$, we have the well established result \cite{Mandel}
\begin{equation}\label{intcorr}
\frac{\langle\hat{I}_1(x') \hat{I}_2(x)\rangle}{\langle\hat{I}_1(x')\rangle\langle\hat{I}_2(x)\rangle}-1
=\left\vert\mbox{Jinc}\left(\frac{kA}{z}|x-x'|\right)\right\vert^2\,.
\end{equation}

\section{Results}
\begin{figure}[hb]
\centering
\subfloat[]{\includegraphics[width=6cm]{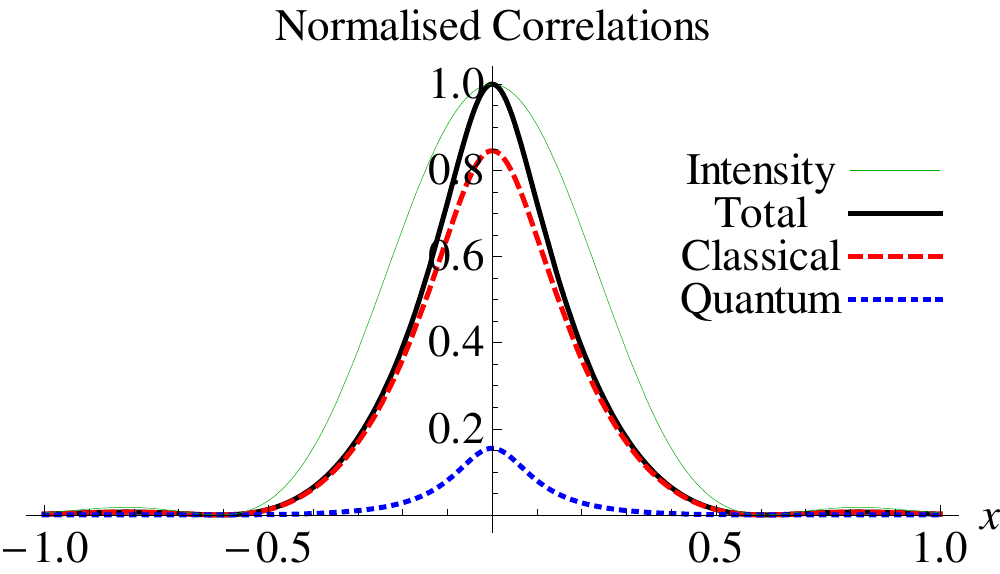}}~
\subfloat[]{\includegraphics[width=6cm]{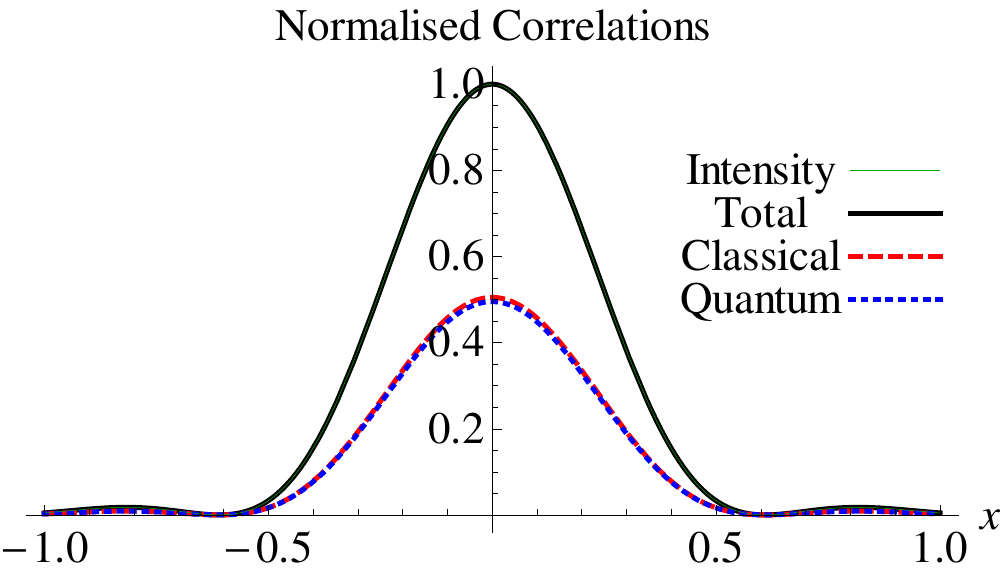}}\\
\subfloat[]{\includegraphics[width=6cm]{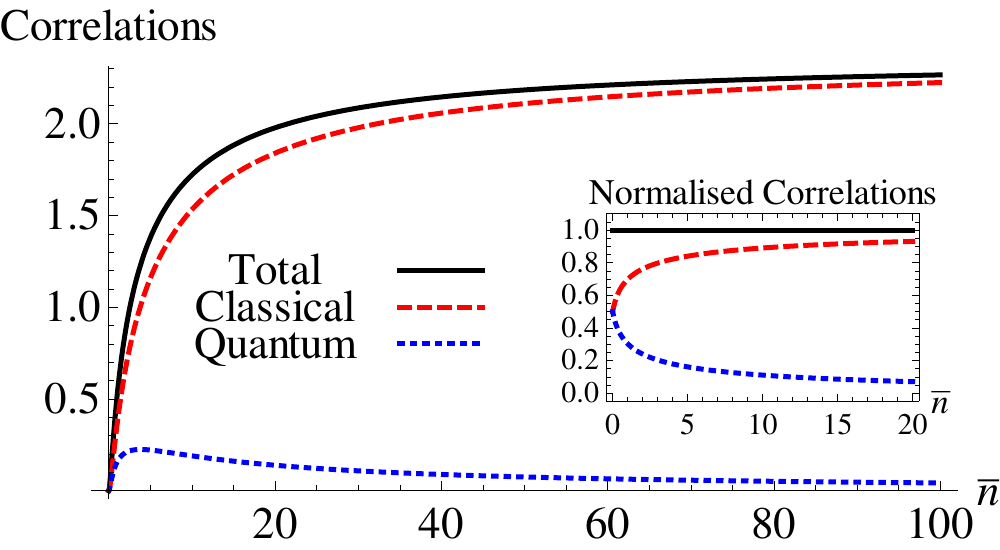}}
\caption{\footnotesize{Plots of the normalised total ($\cal I$, solid black line), classical (${\cal J}$, dashed red line), and quantum (${\cal D}$, dotted blue line) correlations as a function of $x$ for  (a) $\bar{n}=10$ and (b) $\bar{n}=0.01$; the thin green line represents the normalised intensity correlations, which practically coincides with the mutual information in (b). Panel (c) depicts the correlations as a function of $\bar n$ with fixed $x=0$; the inset shows the same correlations normalised by the mutual information, to better highlight the quantum and classical contributions to the total correlations. In all the plots $kA/z=1000$.}}
\label{Graphs}
\end{figure}

To begin with, it is interesting to see how the classical and quantum correlations behave as we scan the detector for fixed values of $\bar{n}$. In all the following results, we have normalised the total correlations to a maximum value of 1 by plotting $\frac{\mathcal{I}(x)}{\mathcal{I}(0)}$; and similarly for quantum and classical ones. For large $\bar{n}$ as in figure~\ref{Graphs}(a), it is apparent that the quantum correlations decay faster than the classical ones and also contribute a much smaller part of the total correlations even at their peak. However, for very small values of $\bar{n}$ as in figure~\ref{Graphs}(b) this is no longer the case: we see that both classical and quantum correlations degrade at the same rate and are approximately equal.
This result stems from the fact that the purity of a Gaussian state relates only to the number of thermal photons \cite{Paris2} present and for small photons counts, we obtain high purity.  Recall that for pure states, $\mathcal{D}=\mathcal{J}=\frac{1}{2}\mathcal{I}$.

To better understand the behaviour of the correlations as we vary $\bar{n}$, it is useful to plot changes against $\bar{n}$ at fixed $x$, see Fig.~\ref{Graphs}(c). It is apparent that the total correlations are composed of a much greater portion of classical correlations than quantum correlations at high photon-count per mode. This gels closely with our intuition: in the regime of high brightness, the quantum component of correlations becomes very small. It is thus easy to see why in certain regimes, the effect is simple to explain classically: it appears that high photon counts wash-out quantumness, increasing the mixedness and simultaneously quashing the presence of discord.

Returning to the regime of high purity (and low photon-count), we find a notable result. The total correlations as quantified by $\mathcal{I}$ match almost exactly the normalised intensity correlations of equation (\ref{intcorr}). Notice once again that classical and quantum correlations (discord) both contribute in equal halves to such total correlations, and thus to the manifestation of the HBT effect itself in this regime.
Very remarkably, for \textit{any} form of cross-correlation of narrowband, thermal light, the correspondence between mutual information and normalised intensity correlations holds analytically in the weak-light regime (up to the third order in $\bar{n}$), as proven in the Appendix.  This provides an intriguing connection and a convenient short-cut for calculating and/or measuring experimentally these correlations.

\section{Discussion and conclusions}
We have provided an example of how the R\'enyi-$2$ entropy \cite{Renyi} can serve as a powerful tool for analysing the behaviour of correlations in the Gaussian world. The presented results, visually summarised in Fig.~\ref{Graphs}, encapsulate not only information about the correlations but also provide, at a mere glance, information about the purity. They also reveal, significantly, that in low-illumination conditions the normalised mutual information can be properly approximated by the normalised intensity correlation.  A somehow similar result has been recently found in the context of thermal light ghost imaging, whereby a properly defined coarse-grained mutual information has been found to match the signal-to-noise ratio of the scheme in the limit of low illumination \cite{Ourpaper}.

Applying the tools of quantum information to the HBT effect clarifies the source of the disagreement in its interpretation and unveils a quantifier of quantumness which behaves particularly intuitively. We have shown that in the limit of a high photon-count per mode the correlations are mostly classical, which matches observations made as far back as the 1950s that the classical `wave-like' features of light emerge in this regime \cite{Gabor}.

We find that for very low photon-counts, however, the quantum correlations are no longer negligible and it was in this regime where the effect originally caused ripples of controversy. This provides a significant example of an insufficiency of following a single definition of classicality. Whilst Scarcelli et al. interpreted their finding of non-local two-photon interference as contradicting previous semi-classical interpretations \cite{Scarcelli}, others have shown this to be incorrect \cite{Gatti2}. The actuality of the situation, and the source of confusion, is that the P-classicality criterion is simply unable to resolve the quantumness in the system.

On the other hand the present discord analysis \textit{does} support the finding that there is a significant presence of quantum effects in the light correlations at low illuminations. As previously mentioned, the strong discrepancy between both definitions of classicality has been recently delineated on rigorous footings \cite{Paris}. It is sufficient for our purposes though to note that the P-classicality is equivalent to separability for bipartite (squeezed) thermal states \cite{Slater} while discord captures all quantum correlations beyond entanglement (in fact, any two correlated Gaussian states will have non-zero discord \cite{QvC}).

In summary, we have utilised a definition of `quantumness' of correlations based on R\'enyi-$2$ Gaussian discord \cite{Renyi,QvC}, which is found to vary continuously from its largest relative value in the low-illumination, high-purity regime, shrinking as the illumination is increased. We believe that this is a naturally relevant way of looking at classicality for such interferometry experiments since the P-classicality reveals very little information and holds even in the seemingly non-classical regime of very few photons.

{\bf Acknowledgments}. We acknowledge the University of Nottingham for financial support under an Early Career Research and Knowledge Transfer Award.

\appendix
\section{}
Here we prove that for two-mode Gaussian state with covariance matrix of the form (\ref{Covmat}) with
$a=b=1+2\bar{n}$ and $c=d= 2\bar{n}f(x)$,
we have when $\bar{n}\ll 1$ that $\frac{\mathcal{I}(x)}{\mathcal{I}(0)} \approx f(x)=\left(\frac{\langle\hat{b}_1^\dagger(x) \hat{b}_2(x')\rangle}{[\langle \hat{I}(x) \rangle \langle\hat{I}(x') \rangle]^{1/2}}\right)^2=\frac{\langle\hat{I}_1(x') \hat{I}_2(x)\rangle}{\langle\hat{I}_1(x')\rangle\langle\hat{I}_2(x)\rangle}-1$.
The rightmost equality is well known to hold for thermal light \cite{Mandel}.
It is easy to show that $\left(\frac{\langle\hat{b}_1^\dagger(x) \hat{b}_2(x')\rangle}{[\langle \hat{I}(x) \rangle \langle\hat{I}(x') \rangle]^{1/2}}\right)^2=f(x)$ arises naturally from the fact that $\langle\hat{b}^\dagger_1(x)\hat{b}_2(x')\rangle=\bar{n}f(x)$ and $\langle\hat{b}^\dagger_i(x) \hat{b}_i(x)\rangle=\bar{n}$ where $i=1,2$.
In order to show that $\frac{\mathcal{I}(x)}{\mathcal{I}(0)} \approx f(x)$, it is useful first to recall that in standard form we can write $\mathcal{I}(x)=\ln\frac{a^2}{a^2-c(x)^2}$. If we write $g(x)=\frac{a^2}{a^2-c(x)^2}$, then we can say $\frac{\mathcal{I}(x)}{\mathcal{I}(0)}=\frac{\ln g(x)}{\ln g(0)}=\log_{g(0)}g(x)$.
This then reduces the problem of showing that $\frac{\mathcal{I}(x)}{\mathcal{I}(0)} \approx f(x)$ to showing that $g(x) \approx g(0)^{f(x)}$. The Taylor expansions of $\frac{(1+2\bar{n})^2}{(1+2\bar{n})^2-(2\bar{n}f(x))^2}$ and of $[\frac{(1+2\bar{n})^2}{(1+2\bar{n})^2-(2\bar{n})^2}]^{f(x)}$ match up to third order in $\bar{n}$ (where we have imposed the condition $f(0)=1$ as a natural consequence of the beamsplitter transformation).
This provides a proof that in low-light source conditions (i.e. small $\bar{n}$), the normalised intensity correlations are almost exactly equal to the normalised R\'enyi-$2$ mutual information in our setting.

\end{document}